\documentclass[conference]{IEEEtran}
\IEEEoverridecommandlockouts
\usepackage{soul}
\usepackage[mathscr]{eucal}
\usepackage[cmex10]{amsmath}
 \usepackage{amssymb,amsmath,amsthm,amsfonts,latexsym}
\usepackage{amsmath,graphicx,bm,xcolor,url,overpic}
\usepackage[caption=false]{subfig} 
\usepackage{array}%array and tabular environments
\usepackage{verbatim}
\usepackage{bm}

\usepackage{verbatim}
\usepackage{textcomp}
\usepackage{mathrsfs}
\usepackage{epstopdf}
\usepackage{cite}
\usepackage{graphicx}
\usepackage{textcomp}
\usepackage{xcolor}
\usepackage[ruled]{algorithm2e}
\newcommand{\openone}{\leavevmode\hbox{\small1\normalsize\kern-.33em1}}

\catcode`~=11 \def\UrlSpecials{\do\~{\kern -.15em\lower .7ex\hbox{~}\kern .04em}} \catcode`~=13 

\allowdisplaybreaks[4]

\newcommand{\nn}{\nonumber}

\newcommand{\calO}{\mathcal{O}}

\newcommand{\rmd}{\mathrm{d}}

\newcommand{\rmI}{\mathrm{I}}

\newcommand{\rmO}{\mathrm{O}}

\newcommand{\rmP}{\mathrm{P}}

\newcommand{\rmR}{\mathrm{R}}
\newcommand{\rms}{\mathrm{s}}

\DeclareMathAlphabet{\mathbsf}{OT1}{cmss}{bx}{n}
\DeclareMathAlphabet{\mathssf}{OT1}{cmss}{m}{sl}% slanted sans serif

\DeclareSymbolFont{bsfletters}{OT1}{cmss}{bx}{n}  
\DeclareSymbolFont{ssfletters}{OT1}{cmss}{m}{n}
\DeclareMathSymbol{\bsfGamma}{0}{bsfletters}{'000}
\DeclareMathSymbol{\ssfGamma}{0}{ssfletters}{'000}
\DeclareMathSymbol{\bsfDelta}{0}{bsfletters}{'001}
\DeclareMathSymbol{\ssfDelta}{0}{ssfletters}{'001}
\DeclareMathSymbol{\bsfTheta}{0}{bsfletters}{'002}
\DeclareMathSymbol{\ssfTheta}{0}{ssfletters}{'002}
\DeclareMathSymbol{\bsfLambda}{0}{bsfletters}{'003}
\DeclareMathSymbol{\ssfLambda}{0}{ssfletters}{'003}
\DeclareMathSymbol{\bsfXi}{0}{bsfletters}{'004}
\DeclareMathSymbol{\ssfXi}{0}{ssfletters}{'004}
\DeclareMathSymbol{\bsfPi}{0}{bsfletters}{'005}
\DeclareMathSymbol{\ssfPi}{0}{ssfletters}{'005}
\DeclareMathSymbol{\bsfSigma}{0}{bsfletters}{'006}
\DeclareMathSymbol{\ssfSigma}{0}{ssfletters}{'006}
\DeclareMathSymbol{\bsfUpsilon}{0}{bsfletters}{'007}
\DeclareMathSymbol{\ssfUpsilon}{0}{ssfletters}{'007}
\DeclareMathSymbol{\bsfPhi}{0}{bsfletters}{'010}
\DeclareMathSymbol{\ssfPhi}{0}{ssfletters}{'010}
\DeclareMathSymbol{\bsfPsi}{0}{bsfletters}{'011}
\DeclareMathSymbol{\ssfPsi}{0}{ssfletters}{'011}
\DeclareMathSymbol{\bsfOmega}{0}{bsfletters}{'012}
\DeclareMathSymbol{\ssfOmega}{0}{ssfletters}{'012}

\usepackage{setspace}
\usepackage{tikz}
\begin{document}
\title{MIMO Radar Meets Polarization-Reconfigurable Antennas: A BCRB Perspective}
\author{\IEEEauthorblockN{Jinpeng Xu and Shuowen Zhang}
	\IEEEauthorblockA{ Department of Electrical and Electronic Engineering, The Hong Kong Polytechnic University\\E-mails:  \{jinpeng.xu, shuowen.zhang\}@polyu.edu.hk }
\thanks{This work was supported in part by the National Natural Science Foundation of China (NSFC) under Grant 62471421, in part by the Hong Kong Research Grants Council (RGC) General Research Fund under Grant 15230022, and in part by the Hong Kong RGC Young Collaborative Research Grant (YCRG) under Grant PolyU C5002-23Y.}}

\maketitle
\begin{abstract}
In this paper, we investigate a novel multiple-input multiple-output (MIMO) radar system aided by phase shifter based polarization-reconfigurable antennas (PRAs). Specifically, a base station (BS) equipped with multiple PRAs at both the transmitter and the receiver aims to sense the \emph{unknown} and \emph{random} angular location parameter of a point target via sending wireless signals and processing the received echo signals reflected by the target, where only \emph{prior distribution information} about the location parameter is available for exploitation. Firstly, we characterize the sensing performance of this novel PRA-based MIMO radar system by deriving the \emph{Bayesian Cram\'er-Rao bound (BCRB)} of the mean-squared error (MSE) in estimating the desired location parameter with prior distribution information. Then, to fully exploit the new design degrees-of-freedom (DoF) empowered by PRAs, we study the joint optimization of the transmit sample covariance matrix as well as the transmit and receive phase shift vectors to minimize the sensing BCRB subject to a transmit power constraint. This problem is non-convex and difficult to solve due to the coupling among optimization variables. To resolve this issue, we develop an alternating optimization (AO) based algorithm which iteratively obtains the \emph{closed-form optimal solution} to each variable with the others being fixed at each time, thus being guaranteed to converge to at least a \emph{stationary point} of the joint optimization problem. Numerical results validate the effectiveness of the proposed algorithm. 
\end{abstract} 
%\begin{IEEEkeywords}
%Polarization-reconfigurable antennas (PRAs), Bayesian Cram\'er-Rao bound (BCRB), multiple-input multiple-output (MIMO) radar, optimization.
%\end{IEEEkeywords}
\vspace{-1mm}
\section{Introduction}
\vspace{-1mm}
Multiple-input multiple-output (MIMO) radar has become a cornerstone technology in modern sensing, owing to its ability to exploit spatial degrees of freedom for performance enhancement\cite{haimovich2008mimo}. By simultaneously transmitting and receiving multiple probing signals via antenna arrays, MIMO radar systems achieve higher parameter identifiability, improved spatial resolution, and robust detection capabilities compared to phased array radar systems\cite{li2008mimo}. The integration of MIMO radar and MIMO communication has also attracted significant research interests in  integrated sensing and communication. 

\begin{figure}[!t]
	\centering
	\includegraphics[width=0.97\linewidth]{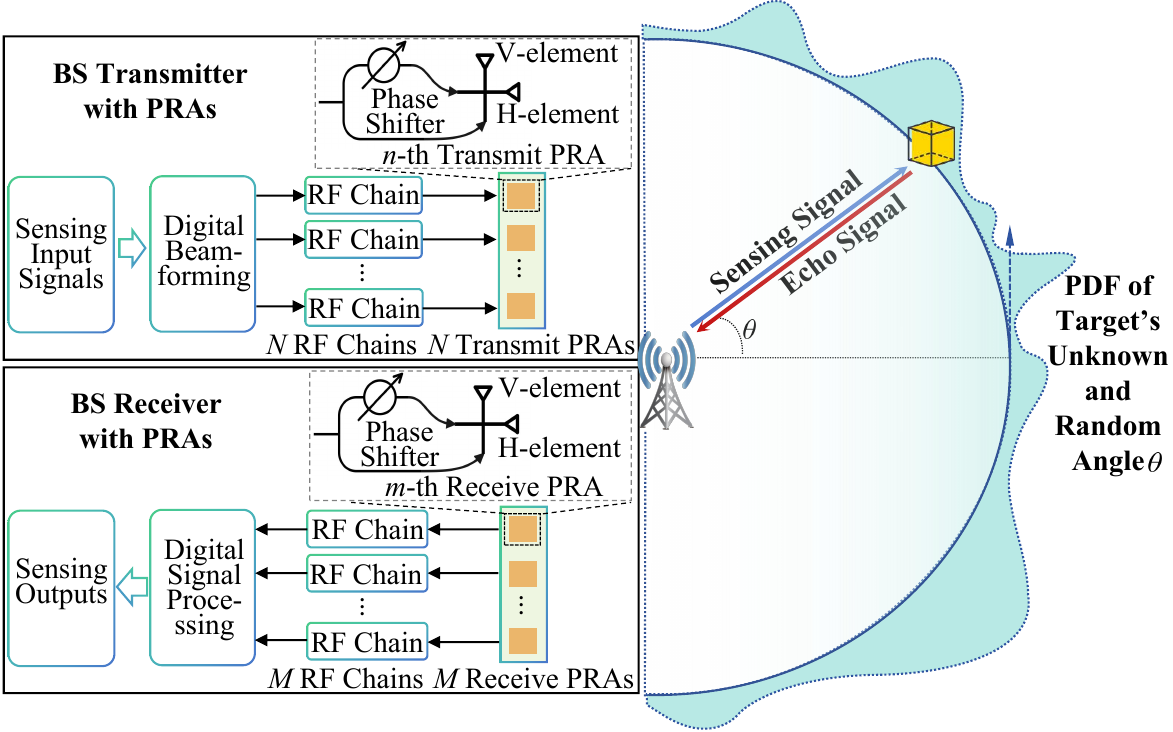}	\vspace{-3mm}
	\caption{Illustration of a MIMO radar system with phase shifter based PRAs.}
	\label{fig:PRA_Sensing} 
	\vspace{-3mm}
\end{figure}
Despite substantial progress, further pushing the performance boundaries of conventional MIMO radar remains challenging, since the spatial domain has been extensively exploited and additional gains from waveform or array design are becoming marginal\cite{ding2025polarforming}. Against this backdrop, polarization, a fundamental property of electromagnetic waves describing the orientation of the electric field, offers a largely untapped resource for radar signal processing\cite{andrews2001tripling}. Polarization offers an extra degree of freedom to enhance sensing performance, especially in challenging multipath and depolarized environments \cite{shahzadi2024dual}. In recent years, interest has emerged in using polarized antennas in MIMO radar to harness this dimension \cite{gogineni2009polarimetric}. Dual- and fully-polarized antenna arrays, each polarization mode connected to a dedicated radio frequency (RF) chain, have shown performance gains by enabling polarization diversity and improving robustness to channel depolarization \cite{kong2018compact}. However, this approach substantially increases both hardware complexity and implementation costs, as each polarization channel requires a dedicated RF chain, which is often prohibitive for large-scale or low-cost radar deployments. Motivated by the need for low-complexity yet flexible polarization control, the recent study \cite{zhou2024polarforming} proposed phase shifter-based polarization-reconfigurable antennas (PRAs) for wireless communications. Distinct from traditional dual-polarized arrays, the proposed approach uses only one RF chain per antenna, with a phase shifter enabling continuous and dynamic polarization control. Specifically, by adjusting the phase difference between orthogonal elements in each PRA, the system can synthesize desired polarization states, including linear, circular, and elliptical. This paradigm, referred to as polarforming\cite{zhou2024polarforming,zhou2025polarforming}, allows the transceiver to match the polarization of its transmit and receive beams to the propagation environment, thereby unlocking the full potential of polarization diversity without incurring excessive cost. However, to the best of our knowledge, there has been no prior studies on the optimization of PRA-aided MIMO radar systems, which motivates our investigation in this work.

On the other hand, in MIMO radar, Cram\'er-Rao bound (CRB) \cite{li2007range} has been widely adopted as a performance metric for lower bounding the sensing mean-squared error (MSE), which is tight in moderate-to-high signal-to-noise ratio (SNR) regimes. However, CRB is a function of the actual values of the parameters to be sensed, which are generally not known before the radar sensing mission is carried out. Therefore, it may not be suitable for guiding the transmit signal design in MIMO radar prior to signal transmission. Although various other performance metrics have been proposed, such as similarity with a desired beampattern \cite{stoica2007probing} and mutual information \cite{yang2007mimo}, their relationships with the MSE are generally implicit. Recently, motivated by the fact that the \emph{distribution information} about the parameters to be sensed can be practically known \emph{a priori} based on target movement pattern or historic data, \emph{posterior Cram\'er-Rao bound (PCRB)} or \emph{Bayesian CRB (BCRB)} has been proposed as a sensing performance metric when such prior distribution information can be leveraged \cite{xu2023mimo,Xu_PCRB_ISAC,yao2025optimal,wang2025hybrid,Kaiyue_JSAC,zhengshuo_TCCN}. Different from CRB, BCRB is only dependent on the parameters' distributions instead of their actual values, thus being suitable for guiding the transmit signal design. 

In this paper, we will conduct an optimization study of a PRA-aided MIMO radar system which aims to sense the \emph{unknown} location parameter of a point target when only its prior distribution information is available for exploitation. We first derive the BCRB of the MSE for estimating the desired parameter, based on which we formulate the joint optimization problem of the transmit sample covariance matrix and the transmit/receive phase shift vectors for reconfiguring PRAs towards BCRB minimization. Although the problem is non-convex and difficult to solve, we propose an efficient alternating optimization (AO) based algorithm which iteratively obtains the closed-form optimal solution to each variable with the other variables being fixed at each time. The proposed algorithm is guaranteed to converge to at least a stationary point of the problem. Numerical results show the superiority of the proposed algorithm over various benchmark schemes.
\vspace{-2mm}
\section{System Model}
\vspace{-2mm}
We consider a MIMO radar system where a multi-antenna base station (BS) equipped with $N\geq 1$ transmit PRAs and $M\geq 1$ co-located receive PRAs aims to sense the \emph{unknown} angular location of a point target denoted by $\theta$ based on the target-reflected received signals at the BS and the \emph{probability density function (PDF)} of $\theta$ denoted by $p_\Theta(\theta)$, as illustrated in Fig. \ref{fig:PRA_Sensing}. In the following, we present the PRA model based on phase shifters and the MIMO radar model.
\vspace{-2mm}
\subsection{PRA Model based on Phase Shifters}
\vspace{-1mm}
We assume that both the BS transmitter and the BS receiver are equipped with a uniform linear array (ULA) of PRAs based on phase shifters. Each PRA is connected to a dedicated RF chain and consists of two orthogonal elements, the V-element for vertical polarization and the H-element for horizontal polarization. A phase shifter, tunable over $[0, 2\pi)$, adjusts the phase difference between the two elements to enable synthesis of the desired polarization states, including linear, circular, and elliptical.\footnote{Due to the use of a phase shifter, the V and H polarization components have equal amplitudes. As a result, both linear and elliptical polarizations have their major axes at $\pm 45^\circ$. For elliptical polarization, the axial ratio is continuously tunable via the phase difference.} Let $\xi_n\in [0, 2\pi)$ denote the phase shift of the $n$-th transmit PRA and $\varphi_m\in [0, 2\pi)$ denote the phase shift of the $m$-th receive PRA. For ease of exposition, we define ${\bm \xi} =[ \xi_1, \cdots, \xi_N ]^T$ and ${\bm \varphi} =[ \varphi_1, \cdots, \varphi_M ]^T$ as phase shift vectors (PSVs) for the transmitter and the receiver, respectively. The polarforming vectors (PFVs) for characterizing the polarization of each $n$-th PRA at the transmitter or each $m$-th PRA at the receiver are given by \cite{zhou2024polarforming} 
\begin{align} \label{PFV_F}
	&{\bm f}(\xi_{n}) \triangleq \frac{1}{\sqrt{2}} \begin{bmatrix} 1,e^{j\xi_{n}} \end{bmatrix}^T,~\xi_{n}\in[0,2\pi),\ n=1,\cdots,N,\\
	&{\bm e}(\varphi_{m}) \triangleq \begin{bmatrix} 1,e^{j\varphi_{m}}\end{bmatrix}^T, ~\varphi_{m}\in[0,2\pi),\ m=1,\cdots,M.\label{PFV_E}
\end{align}
The factor $\frac{1}{\sqrt{2}}$ normalizes the signal to meet transmit power constraint.
Furthermore, the polarforming matrices (PFMs) at the transmitter and the receiver are modeled as follows by aggregating individual PFVs into block-diagonal matrices:
\begin{align}
	{\bm F}({\bm \xi}) &\triangleq {\rm blkdiag}\left\{ {\bm f}(\xi_{1}),\cdots, {\bm f}(\xi_{N}) \right\} \in \mathbb{C}^{2N\times N}, \\
	{\bm E}({\bm \varphi}) &\triangleq {\rm blkdiag}\left\{{\bm e}(\varphi_1), \cdots,{\bm e}(\varphi_M) \right\}\in \mathbb{C}^{2M\times M}.
\end{align}
\subsection{MIMO Radar Model}
\vspace{-2mm}
Let $L \geq 1$ denote the number of transmit probing samples for estimating $\theta$ in the MIMO radar system. Let $\bm{x}_l \in \mathbb{C}^{N \times 1}$ denote the baseband probing signal vector in the $l$-th sample interval from the $N$ RF chains. Denote $\bm{X} = [\bm{x}_1, \ldots, \bm{x}_L]$ as the collection of probing signals, for which the sample covariance matrix is given by $\bm{R}_X=\frac{1}{L}\sum_{l=1}^L\bm{x}_l\bm{x}_l^H =\frac{1}{L}\bm{XX}^H$. Let $P$ denote the total transmit power constraint, which yields $\mathrm{tr}(\bm{R}_X)\leq P$. After transmit polarforming, the signals are reflected by the target and received back at the BS receiver. 

For the purpose of revealing insights, we consider a line-of-sight (LoS) channel model between the BS and the target. Let $\bm a(\theta)\in \mathbb{C}^{N\times 1}$ and $\bm b(\theta)\in \mathbb{C}^{M\times 1}$ denote the transmit and receive array steering vectors, respectively. Each element in $\bm a(\theta)$ or $\bm b(\theta)$ is given by $a_n(\theta)=e^{-j\frac{\pi}{\lambda}d(N-2n+1)\sin\theta}, n=1,\cdots,N$, or $b_m(\theta)=e^{-j\frac{\pi}{\lambda}d(M-2m+1)\sin\theta}, m=1,\cdots,M$, respectively, where $d$ denotes the antenna spacing and $\lambda$ denotes the wavelength. Let $\psi\in \mathbb{C}$ denote the radar cross-section (RCS) coefficient at the target. The round-trip channel gain is given by $\frac{\beta_0}{r^2}$, where $\beta_0$ denotes the reference channel power at $1$ meter (m), and $r$ denotes the BS-target distance in m. Define $\alpha \overset{\Delta}{=}\frac{\beta_0}{r^2}\psi = \alpha_\rmR + j\alpha_\rmI$ as the overall reflection coefficient between the BS transmitter and the BS receiver, which is considered as an \emph{unknown} parameter with known PDF. The overall propagation channel from the BS transmitter to the BS receiver via target reflection is thus given by
\begin{align}\label{channel}
	{{\bm H}}(\theta,\alpha)=\alpha \bm b(\theta)\bm a^H(\theta)\in \mathbb{C}^{M\times N }.
\end{align}
Let ${\bm \Psi}$ denote the coupling and crosstalk characteristics between transmit and receive polarization modeled as
\begin{equation}
	{\bm \Psi} = \frac{1}{\sqrt{1+\chi}} \begin{bmatrix} 1 & \sqrt{\chi}  \\  \sqrt{\chi} &1 \end{bmatrix},
\end{equation}
where $\chi$ represents the inverse cross-polarization discrimination (XPD) and indicates the degree of channel depolarization.\footnote{In LoS scenario, coupling and crosstalk arise from finite antenna isolation and array orientation\cite{Oestges_dualpolar}; for simplicity, both effects are subsumed into ${\bm \Psi} $.} According to \cite{song2013millimeter}, the polarized overall channel matrix from the BS transmitter to the BS receiver is given by
\begin{align} \label{PCM}
	\bm{P}(\theta,\alpha) &= \begin{bmatrix}
		{\bm P}_{11}(\theta,\alpha) & \cdots & {\bm P}_{1N}(\theta,\alpha) \\
		\vdots & \ddots & \vdots \\
		{\bm P}_{M1}(\theta,\alpha) & \cdots & {\bm P}_{MN}(\theta,\alpha)
	\end{bmatrix} \\
	&= {\bm \Psi} \otimes { {\bm H}}(\theta,\alpha)\in \mathbb{C}^{2M \times 2N},
\end{align}
where $\otimes$ denotes the Kronecker product.

With polarforming at the BS transmitter and receiver, the received signal vector at the $l$-th sample is given by
\begin{align}
	\bm{y}_l={\bm E}^H({\bm \varphi}) {\bm P}(\theta,\alpha) {\bm F}({\bm \xi})\bm{x}_l+\bm{n}_l,\ l=1,\cdots,L,
\end{align}
where $\bm{n}_l\sim\mathcal{CN}(0,\sigma_\rms^2\bm{I}_{M})$ denotes the circularly symmetric complex Gaussian (CSCG) noise vector, with $\sigma_\rms^2$ denoting the average noise power. Define $	{\bm G}(\theta,\alpha,{\bm \xi}, {\bm \varphi}) \triangleq {\bm E}^H({\bm \varphi}) {\bm P}(\theta,\alpha) {\bm F}({\bm \xi})$ as the ``\emph{effective channel matrix}'' from the BS transmit RF chains and the BS receiver RF chains, which is determined by the target's parameters $\theta$, $\alpha$, and the transmit/receive polarforming characterized by the phase shifts in $\bm\xi$ and $\bm \varphi$. Denote $\bm N=[\bm{n}_1,...,\bm{n}_L]$. The collection of received signal vectors over $L$ samples is thus given by
\begin{align}\label{Y}
	\bm{Y}=[\bm{y}_1,...,\bm{y}_L]={\bm G}(\theta,\alpha,{\bm \xi}, {\bm \varphi})\bm{X}+\bm N.
\end{align}

Since both the reflection coefficient $\alpha$ and $\theta$ are unknown in $\bm{Y}$, they need to be jointly estimated to obtain an accurate estimate of $\theta$, based on both the received signals in $\bm{Y}$ as well as the known PDFs of $\theta$ and $\alpha$. In the following, we characterize the estimation performance of $\theta$.
 \vspace{-4mm}
\section{Characterization of Angular Location Estimation Performance via BCRB}\label{derivative_BCRB}
\vspace{-4mm}
In this section, we aim to characterize the estimation performance of the target's angular location $\theta$ via the PRA-aided MIMO radar system with prior distribution information. Note that the MSE is generally difficult to be quantified analytically, especially in the considered system with complex PRA. Therefore, we aim to characterize a lower bound of MSE via deriving the BCRB, which will become increasingly tight as the SNR increases.

Specifically, let $\bm{\zeta}\!\!=\!\![\theta,\alpha_\rmR,\alpha_\rmI]^T$ denote the collection of all the unknown components in $\theta$ and $\alpha$ which need to be jointly estimated to obtain an accurate estimate of $\theta$. We assume that all the entries in $\bm{\zeta}$ are independent of each other, and let $p_{\alpha_\rmR}(\alpha_\rmR)$ and $p_{\alpha_\rmI}(\alpha_\rmI)$ denote the PDFs of $\alpha_\rmR$ and $\alpha_\rmI$, respectively. The joint PDF of $\bm{\zeta}$ is thus given by $p_Z(\bm{\zeta})=p_\Theta(\theta)p_{\alpha_\rmR}(\alpha_\rmR)p_{\alpha_\rmI}(\alpha_\rmI)$. Furthermore, we consider a mild condition of  $\int\alpha_{\mathrm{R}}p_{\alpha_{\mathrm{R}}}(\alpha_{\mathrm{R}})\rmd\alpha_{\mathrm{R}}\!=\!\int\alpha_{\mathrm{I}}p_{\alpha_{\mathrm{I}}}(\alpha_{\mathrm{I}})\rmd\alpha_{\mathrm{I}}\!=\!0$\cite{wang2025hybrid}, which holds for a vast variety of distributions (e.g., when the BS-target distance follows the uniform distribution, and the RCS coefficient $\psi$ follows the CSCG distribution).

The Bayesian Fisher information matrix (BFIM) for estimating $\boldsymbol{\zeta}$ is given by \cite{Van_Trees_PCRB}
\begin{equation}\label{FIM}
	\bm{J} = \bm{J}_\rmO +\bm{J}_\rmP.
\end{equation}
Specifically, $\bm{J}_\rmO\in \mathbb{R}^{3\times 3}$ denotes the BFIM from \emph{observation}, i.e., $\bm{Y}$. Let $f(\bm{Y}|\boldsymbol{\zeta})$ denote the conditional PDF of $\bm{Y}$ with given $\bm{\zeta}$. $\bm{J}_\rmO$ is then given by \cite{Van_Trees_PCRB}
\begin{align}\label{Fo_pf}
	\bm{J}_\rmO=&\mathbb{E}_{\bm{Y},\boldsymbol{\zeta} }\left[\frac{\partial \ln(f(\bm{Y}|\boldsymbol{\zeta}))}{\partial \boldsymbol{\zeta} }\left(\frac{\partial \ln(f(\bm{Y}|\boldsymbol{\zeta}))}{\partial \boldsymbol{\zeta} }\right)^H\right]\\
	=&\left[
	\begin{array}{ll}
		&{ J}_\rmO^{\theta\theta}  \qquad          \bm{J}_\rmO^{\theta\alpha}     \\
		&\bm{J}_\rmO^{{\theta\alpha}^H}\quad  \bm{J}_\rmO^{\alpha\alpha}
	\end{array}
	\right],
\end{align}
where the log-likelihood function for estimating $\bm{\zeta}$ from the observations in $\bm{Y}$ can be derived as
\begin{align}
	&\ln(f(\bm{Y}|\boldsymbol{\zeta}))=\frac{2}{\sigma_\rms^2}\Re\{ \mathrm{tr}(\bm{X}^H{\bm G}^H(\theta,\alpha,  \bm\xi, \bm\varphi)\bm{Y })\}\nn\\
	&-\frac{\|\bm{Y }\|_F^2 + \|\bm{G} (\theta,\alpha,  \bm\xi, \bm\varphi)\bm{X}\|_F^2 }{\sigma_\rms^2}-ML\ln(\pi\sigma_\rms^2).
\end{align}
We further define $\tilde{\bm A}_1\overset{\Delta}{=}\int\left(\! \dot{\bm{b}}(\theta)\bm{a}^H \!(\theta)\!+\!\bm{b}(\theta)\dot{\bm{a}}^H\! (\theta)\!\right)\!p_\Theta(\theta)\rmd\theta$, $\tilde{\bm A}_2\overset{\Delta}{=}\int \left( \bm{b}(\theta)\bm{a}(\theta)^H \right)p_\Theta(\theta)\rmd\theta$. Based on this, each block in $\bm{J}_\rmO$ can be expressed as
\begin{align}
	{ J}_\rmO^{\theta\theta} &\!=\! \frac{2L}{\sigma_\rms^2}      \mathbb{E}_{\boldsymbol{\zeta} }\!\left[  \left(\alpha_\rmR^2+\alpha_\rmI^2\right) \mathrm{tr}\! \left( \bm Q(\bm{\xi},\bm{\varphi})\bm R_X \bm Q^H\!(\bm{\xi},\bm{\varphi})\right)\right]\\
	&= \frac{2L\gamma}{\sigma_\rms^2} \mathrm{tr} \left( \bm Q(\bm{\xi},\bm{\varphi})\bm R_X \bm Q^H(\bm{\xi},\bm{\varphi})\right),
	\\
	{{\bm J}}_\rmO^{\theta\alpha}   &= \frac{2L}{\sigma_\rms^2}      \mathbb{E}_{\boldsymbol{\zeta} }\left[ \mathrm{tr} \left( \bm Q(\bm{\xi},\bm{\varphi})\bm R_X \bm O^H(\bm{\xi},\bm{\varphi})\right)[\alpha_\rmR,\alpha_\rmI]\right]\\
	&=\bm 0,\\
	{{\bm J}}_\rmO^{\alpha\alpha} &= \frac{2L}{\sigma_\rms^2}      \mathbb{E}_{\boldsymbol{\zeta} }\left[ \mathrm{tr}\left( \bm O(\bm{\xi},\bm{\varphi})\bm R_X \bm O^H(\bm{\xi},\bm{\varphi})\right)\bm I_2 \right]\\
	&=\frac{2L}{\sigma_\rms^2} \mathrm{tr} \left( \bm O(\bm{\xi},\bm{\varphi})\bm R_X \bm O^H(\bm{\xi},\bm{\varphi})\right)\bm I_2 ,
\end{align}
where $\gamma\triangleq\int\alpha_{\mathrm{R}}^2p_{\alpha_{\mathrm{R}}}(\alpha_{\mathrm{R}})\rmd\alpha_{\mathrm{R}}+\int\alpha_{\mathrm{I}}^2p_{\alpha_{\mathrm{I}}}(\alpha_{\mathrm{I}})\rmd\alpha_{\mathrm{I}}$, and 
\begin{align}
	\bm Q(\bm{\xi},\bm{\varphi})=\bm{E}^H (\bm{\varphi}) ({\bm \Psi} \otimes\tilde{\bm A}_1) \bm{F}(\bm{\xi})\in \mathbb{C}^{M\times N },
\end{align}
\begin{align}
	\bm O(\bm{\xi},\bm{\varphi})=\bm{E}^H (\bm{\varphi}) ({\bm \Psi}\otimes{\tilde{\bm A}_2}) \bm{F}(\bm{\xi})\in \mathbb{C}^{M\times N }.
\end{align}
 
On the other hand, $\bm{J}_\rmP\in \mathbb{R}^{3\times 3 }$ denotes the BFIM from \emph{prior distribution information}, which is given by
\begin{align}\label{Fp_pf}
	\bm{J}_\rmP=&\mathbb{E}_{\boldsymbol{\zeta} }\left[\frac{\partial \ln(p_Z(\boldsymbol{\zeta} ))}{\partial \boldsymbol{\zeta} }\left(\frac{\partial \ln(p_Z(\boldsymbol{\zeta} ))}{\partial \boldsymbol{\zeta} }\right)^H\right]\\
	=&\mathrm{diag}\bigg\{\mathbb{E}_\theta\Big[\Big(\frac{\partial\ln(p_{\Theta}(\theta))}{\partial\theta}\Big)^2\Big]
	,\mathbb{E}_{\alpha_{\mathrm{R}}}\Big[\Big(\frac{\partial\ln(p_{\alpha_{\mathrm{R}}}(\alpha_{\mathrm{R}}))}{\partial\alpha_{\mathrm{R}}}\Big)^2\Big],\nonumber\\
	&\mathbb{E}_{\alpha_{\mathrm{I}}}\Big[\Big(\frac{\partial\ln(p_{\alpha_{\mathrm{I}}}(\alpha_{\mathrm{I}}))}{\partial\alpha_{\mathrm{I}}}\Big)^2\Big]\bigg\}.
\end{align} 
The BCRB matrix for the MSE matrix of estimating $\bm{\zeta}$ is given by $\bm{J}^{-1}=(\bm{J}_\rmO +\bm{J}_\rmP)^{-1}$ \cite{Van_Trees_PCRB}. The BCRB for the MSE in estimating the desired parameter $\theta$ is thus given below by further noting that $\bm{J}$ is a block-diagonal matrix:
\begin{align}
&\!\!\!	\mathrm{BCRB}_{\theta}\left(\bm\xi, \bm\varphi,\bm{R}_X\right)=[\bm{J}^{-1}]_{1,1}\nn\\*
	&\!\!	\!=\!\frac{1}{\frac{2L\gamma}{\sigma_\rms^2}\mathrm{tr} \!\left( \bm Q(\bm{\xi},\bm{\varphi})\bm R_X \bm Q^H(\bm{\xi},\bm{\varphi})\right)\!+\!\mathbb{E}_\theta\Big[\!\Big(\frac{\partial\ln(p_{\Theta}(\theta))}{\partial\theta}\Big)^2\Big]}.\label{PCRB}	
\end{align}

Notice that $\mathrm{BCRB}_{\theta}(\bm\xi, \bm\varphi,\bm{R}_X)$ is dependent on both the prior distribution information about $\theta$ and $\alpha$, as well as the PSVs at the BS transmitter and BS receiver, $\bm\xi$, $\bm\varphi$, and the transmit sample covariance matrix $\bm{R}_X$. In the following, we aim to jointly optimize the PSVs and the sample covariance matrix to minimize the BCRB in estimating $\theta$.

\section{Problem Formulation}
Note that minimizing $\mathrm{BCRB}_{\theta}\left(\bm\xi, \bm\varphi,\bm{R}_X\right)$ is equivalent to maximizing $\mathrm{tr}(\bm Q(\bm{\xi},\bm{\varphi})\bm R_X \bm Q^H(\bm{\xi},\bm{\varphi}))$. The optimization problem is thus formulated as
\begin{align} \label{P1-1}
	\mbox{(P1)} \quad  \mathop{\mathrm{max}}_{\bm\xi, \bm\varphi,\bm{R}_X\succeq \bm{0}   } \quad  & \mathrm{tr} \left( \bm Q(\bm{\xi},\bm{\varphi})\bm R_X \bm Q^H(\bm{\xi},\bm{\varphi})\right)
	\\*
	{\mathrm{s.t.}}~~~ \quad &\xi_n\in [0,2\pi),\ n=1,\cdots,N\\
	& \varphi_m\in [0,2\pi),\ m=1,\cdots,M\\
	&\mathrm{tr}(\bm{R}_X)\leq P.\label{power_constraint}
\end{align}
Problem (P1) is challenging to solve due to the complicated coupling between $\bm{R}_X$ and  $\bm Q(\bm{\xi},\bm{\varphi})$, as well as the complex form of $\bm Q(\bm{\xi},\bm{\varphi})$ with respect to $\bm{\xi}$ and $\bm{\varphi}$. To resolve these challenges, we propose an AO-based algorithm that iteratively optimizes one variable at each time with the others being fixed.
\vspace{-5mm}
\addtolength{\topmargin}{0.04in}
\section{Proposed Solution to  Problem (P1)}
In the proposed AO-based algorithm, we first initialize $\bm\xi$ and $\bm\varphi$ (e.g., via randomly generating the phase shifts therein), and then iteratively solve the following sub-problems, each for optimizing $\bm{R}_X$ or one element in $\bm\xi$ or $\bm\varphi$.
\vspace{-2mm}
\subsection{Optimization of $\bm{R}_X$ with Given $\bm\xi$ and $\bm\varphi$}
With given transmit and receive PSVs $\bm\xi$ and $\bm\varphi$, (P1) is reduced to the following sub-problem for optimizing $\bm{R}_X$:
\begin{align} \label{P1-2}
	\mbox{(P1-I)} \quad  \mathop{\mathrm{max}}_{\bm{R}_X \succeq \bm{0}: \eqref{power_constraint}} \quad  & \mathrm{tr} ( \bm Q(\bm{ \xi},\bm{\varphi})\bm R_X \bm Q^H(\bm{\xi},\bm{\varphi}))
\end{align}
(P1-I) is a convex semi-definite program (SDP). Let ${\tilde{\bm Q}}= \bm Q^H(\bm{\xi},\bm{\varphi}) \bm Q(\bm{\xi},\bm{\varphi})$. Note that $\mathrm{tr} ( \bm Q(\bm{ \xi},\bm{\varphi})\bm R_X \bm Q^H(\bm{\xi},\bm{\varphi}))=\mathrm{tr}({\tilde{\bm Q}}\bm R_X)$. Thus, an optimal solution to (P1-I) can be shown to be given by $\boldsymbol{R}^\star_X=P\tilde {\bm{q}}\tilde {\bm{q}}^H$,  where $\tilde {\bm{q}}$ denotes the eigenvector corresponding to the strongest eigenvalue of ${\tilde{\bm Q}}$ \cite{Xu_PCRB_ISAC}. 
\vspace{-2mm}
\subsection{Optimization of $\varphi_m$ with Given $\bm{R}_X$, $\bm\xi$, $\{\varphi_\ell,\ell\neq m\}_{\ell=1}^M$}
Note that $\bm Q(\bm{\xi},\bm{\varphi})$ can be explicitly expressed as
\begin{align}\label{Q_xi_phi}
	&\bm{Q}(\bm{\xi},\bm{\varphi})\!=\\
	&\begin{bmatrix}
		\bm{e}^H(\varphi_1) &  &  \\
		& \ddots  &  \\
		&  & \bm{e}^H(\varphi_M)
	\end{bmatrix}
	{\bm \Psi}\otimes\tilde{\bm A_1}
	\begin{bmatrix}
		\bm{f}(\xi_1) &  &  \\
		& \ddots & \\
		&  & \bm{f}(\xi_N)
	\end{bmatrix}\!.\nonumber
\end{align}
Each element in $\bm Q(\bm{\xi},\bm{\varphi})$ can be further expressed as
\begin{align}\label{entry_Q}
	q_{mn}\! \!\triangleq\![\bm{Q}(\bm{\xi},\bm{\varphi})] _{m,n}\!=\! \bm{e}^H(\varphi_m)	{\bm \Psi}[\tilde{\bm A_1}]_{m,n}\bm{f}(\xi_n), \forall m,n.
\end{align}
Note that $\bm Q(\bm{\xi},\bm{\varphi})=\bm{E}^H(\bm{\varphi}) \bm A_1 \bm{F}(\bm{\xi})$, we further have
\begin{align}
	&[\bm{Q}(\bm{\xi},\bm{\varphi}){\bm{R}_{X}}\bm{Q}^H(\bm{\xi},\bm{\varphi})]_{m,m}=\! \sum_{n=1}^N \sum_{i=1}^N q_{mn} [\bm{R}_{X}]_{n,i} q_{mi}^*\\
	&=\!\bm{e}^H (\varphi_m) \!\Bigg( \sum_{n=1}^N \sum_{i=1}^N \bm{\Psi}[\tilde{\bm{A}}_1]_{m,n}\bm{f}(\xi_n)[\bm{R}_{X}]_{n,i}\bm{f}^H(\xi_i)[\tilde{\bm{A}}_1]_{m,i}^* \nn\\
	&\quad   \times \bm{\Psi}  \big) \bm{e}(\varphi_m).
\end{align}
Hence, the objective function of (P1) is given by
\begin{align} \label{trQRO}
	&\mathrm{tr} \left( \bm Q(\bm{\xi},\bm{\varphi})\bm R_{X} \bm Q^H(\bm{\xi},\bm{\varphi})\right)\nn\\
	& =  \sum_{m=1}^M \sum_{n=1}^N \sum_{i=1}^N q_{mn} [\bm{R}_{X}]_{n,i} q_{mi}^* \\
	&=\sum_{m=1}^M \bm{e}^H(\varphi_m) \left( \sum_{n=1}^N \sum_{i=1}^N \bm{\Psi}[\tilde{\bm{A}}_1]_{m,n} \bm{f}(\xi_n)[\bm{R}_{X}]_{n,i}\bm{f}^H(\xi_i)\right. \nn\\
	&\quad \times[\tilde{\bm{A}}_1]_{m,i}^* \bm{\Psi} \Big) \bm{e} ( \varphi_m)\\
	&=   \bm{e}^H(\varphi_m) \bm{K}_m\bm{e}(\varphi_m) + \sum_{\iota=1,\iota\ne m}^M \bm{e}^H(\varphi_\iota) \bm{K}_\iota\bm{e}(\varphi_\iota).
\end{align}
Then, define
\begin{align}\label{K_m}
	\!\!\bm{K}_m \!\triangleq\!  \!\sum_{n=1}^N \!\sum_{i=1}^N \bm{\Psi}[\tilde{\bm{A}}_1]_{m,n}\bm{f}(\xi_n)[\bm{R}_{X}]_{n,i}\bm{f}^H(\xi_i)[\tilde{\bm{A}}_1]_{m,i}^* \bm{\Psi}.\!\!
\end{align}
The sub-problem for optimizing $\varphi_m$ with given $\bm{R}_X$, $\bm\xi$, $\{\varphi_\ell,\ell\neq m\}_{\ell=1}^M$ is then formulated as
\begin{align} 
	\mbox{(P1-II)} \quad  \mathop{\mathrm{max}}_{\varphi_m\in [0,2\pi)} \quad  & \bm{e}^H(\varphi_m) \bm{K}_m\bm{e}(\varphi_m).
\end{align}
Note that ${\bm K}_m$ is the summation of Hermitian matrices, and $\bm{e}(\varphi_m)=[1,e^{j\varphi_m}]^T$. Therefore, the optimal phase shift $\varphi_m$ for the $m$-th receive PRA can be shown to be given by
\begin{equation} \label{opt_phim}
	\varphi_m^\star = \angle{[{\bm K}_m]_{21}}.
\end{equation}
\vspace{-6mm}
\subsection{Optimization of $\xi_n$ with Given $\bm{R}_X$, $\bm\varphi$, $\{\xi_\ell,\ell\neq n\}_{\ell=1}^N$}
In this subsection, we aim to optimize $\xi_n$ with given $\bm{R}_X$, $\bm\varphi$, $\{\xi_\ell,\ell\neq n\}_{\ell=1}^N$. To this end, we express the objective function of (P1) as
\begin{align} \label{Am}
	&\mathrm{tr} \left( \bm Q(\bm{\xi},\bm{\varphi})\bm R_{X} \bm Q^H(\bm{\xi},\bm{\varphi})\right)\nn\\
	&=\sum_{m=1}^M \sum_{n=1}^N \sum_{i=1}^N q_{mn} [\bm{R}_{X}]_{n,i} q_{mi}^* \\
	&=\sum_{n=1}^N\left( \sum_{m=1}^M  \sum_{i=1}^N q_{mn} [\bm{R}_{X}]_{n,i} q_{mi}^* \right)\\
	&= \sum_{n=1}^N  \bm{f}^H(\xi_n)\left( \sum_{m=1}^M \sum_{i=1}^N \bm{\Psi}[\tilde{\bm{A}}_1]_{m,n}\bm{e}(\varphi_m)[\bm{R}_{X}]_{n,i}\bm{e}^H(\varphi_m)\right.\nn\\
	&\quad \times[\tilde{\bm{A}}_1]_{m,i}^* \bm{\Psi}\Big)  \bm{f}(\xi_n)\\*
	&=   \bm{f}^H(\xi_n) {\bm{V}_n}\bm{f}(\xi_n) + \sum_{\ell=1,\ell\ne n}^N \bm{f}^H(\xi_\ell) {\bm{V}_\ell}\bm{f}(\xi_\ell).
\end{align}
Define
\begin{align}\label{K_n}
	\bm{V}_n \triangleq \sum_{m=1}^M \sum_{i=1}^N \bm{\Psi}[\tilde{\bm{A}}_1]_{m,n}\bm{e}(\varphi_m)[\bm{R}_{X}]_{n,i}\bm{e}^H(\varphi_m)[\tilde{\bm{A}}_1]_{m,i}^* \bm{\Psi}.
\end{align}
The sub-problem for optimizing $\xi_n$ with given $\bm{R}_X$, $\bm\varphi$, $\{\xi_\ell,\ell\neq n\}_{\ell=1}^N$ is then formulated as
\begin{align} 
	\mbox{(P1-III)} \quad  \mathop{\mathrm{max}}_{\xi_n\in [0,2\pi)} \quad  & \bm{f}^H(\xi_n) \bm{V}_n\bm{f}(\xi_n).
\end{align}
Note that ${\bm V}_n$ is the summation of Hermitian matrices, and ${\bm{f}(\xi_n)}=\frac{1}{\sqrt{2}}[1,e^{j\xi_n}]^T$. Therefore, the optimal phase shift $\xi_n$ for the $n$-th transmit PRA can be shown to be given by
\begin{equation} \label{opt_thetan}
	\xi_n^\star = \angle{[{\bm V}_n]_{21}}.
\end{equation}
\subsection{Summary of the Proposed AO-based Algorithm}
The proposed AO-based algorithm is summarized in Algorithm \ref{algo1}. Specifically, the proposed algorithm iteratively obtains the \emph{optimal solutions} to (P1-I), (P1-II) for $m=1,\cdots,M$, and (P1-III) for $n=1,\cdots,N$ in closed forms. Note that $\bm{R}_X$, $\varphi_m$, and $\xi_n$ are not coupled in any constraint of (P1). Therefore, the proposed algorithm is guaranteed to converge to at least a \emph{stationary point} of Problem (P1).
\begin{algorithm}[!t]\label{algo1}
	\caption{Proposed AO-based algorithm for (P1)}
	\textbf{Initialize:} Randomly generate $\bm{\xi}$ and $ \bm{\varphi}$\;
	\Repeat{convergence}{
	%	\textbf{Optimize $\bm{R}_X$ for given $\bm{\xi}^{(t)}, \bm{\varphi}^{(t)}$:}\\
		\Indp
		Compute $\tilde{\bm Q} = \bm Q^H(\bm{\xi},\bm{\varphi}) \bm Q(\bm{\xi},\bm{\varphi})$\;
		Obtain strongest eigenvector {$\tilde{\bm{q}}$} of $\tilde{\bm Q}$\;
		Set $\bm{R}_X = P\, {\tilde{\bm{q}} \tilde{\bm{q}}^H}$\;
		\Indm
		
	%	\textbf{Optimize  receive PSV $\bm \varphi$ for given $\bm{R}_X^{(t+1)}$ :}\\
		\For{$m=1$ \KwTo $M$}{
			Compute matrix $\bm{K}_m$ based on \eqref{K_m}\;
			Set $\varphi_m = \angle \left([\bm{K}_m]_{21}\right)$\;
		}
		
	%	\textbf{Optimize transmit PSV $\bm \xi$ for given $\bm{R}_X^{(t+1)}$:}\\
		\For{$n=1$ \KwTo $N$}{
			Compute matrix $\bm{V}_n$ based on \eqref{K_n}\;
			Set $\xi_n = \angle \left([\bm{V}_n]_{21}\right)$\;
		}
		
	}
	\Return{$\bm{\xi}, \bm{\varphi}, \bm{R}_X$}.
\end{algorithm}
\vspace{-2mm}\addtolength{\topmargin}{-0.02in}
\subsection{Complexity Analysis}
The per-iteration computational complexity of the proposed AO algorithm is dominated by $\calO(N^3)$ for updating the covariance matrix ${\bm R}_X$ via eigenvalue decomposition, $\calO(MN^2)$ for updating receive PSV $\bm{\varphi}$, $\calO(N^2M)$ for updating transmit PSV ${\bm\xi}$, $\calO(MN)$ for constructing the polarforming matrix ${\bm Q(\bm\xi, \bm\varphi)}$, and $\calO(MN^2)$ for evaluating the objective $g(\bm\xi, \bm\varphi, {\bm R}_X)$. Thus, the overall per-iteration complexity is $\calO(N^3 + 3MN^2 + MN)$, leading to a total complexity of $\calO\left( T \left( N^3 + 3MN^2 + MN \right) \right)$ over $T$ iterations.

\vspace{-2mm}
\section{Numerical Results}
In this section, numerical results are provided to evaluate the performance of the proposed PRA-aided MIMO radar system. We set  $N=12$,  $M=12$, $d=\frac{\lambda}{2}$,  $L=25$, and $P=30$ dBm. Specifically, the inverse  XPD is set as $\chi=0.2$\cite{calcev2007wideband}. We set the overall reflection coefficient $\alpha$ as $\alpha \sim \mathcal{CN}(0, 2\sigma_{\alpha}^2)$, where $\sigma_{\alpha}^2 = 10^{-12}$ \cite{wang2025hybrid}. Furthermore, we assume that the PDF of $\theta$ follows a Gaussian mixture model, which is the weighted summation of $K\geq 1$ Gaussian PDFs and is given by $p_{\Theta}(\theta)=\sum_{k=1}^K\frac{p_k}{\sqrt{2\pi}\sigma_k}\exp({-\frac{(\theta-\theta_k)^2}{2\sigma_k^2}})$. Specifically, $\theta_k\in[0, {\pi})$ and $\sigma_k^2$ are the mean and variance of the $k$-th Gaussian PDF, respectively, and $p_k$ denotes the weight that satisfies $\sum_{k=1}^Kp_k=1$. We further set $K=4$; $\theta_1=0.3$, $\theta_2= 1.2$, $\theta_3=2.5$, $\theta_4=2.9 $; $\sigma_1^2=10^{-2}$, $\sigma_2^2=10^{-2}$, $\sigma_3^2=10^{-2}$, $\sigma_4^2=10^{-2}$; $p_1=0.2$, $p_2=0.6$, $p_3=0.1$, $p_4=0.1$. For comparison, the following benchmark schemes for the BS transceiver design are considered:
\begin{itemize}
	\item {\bf Benchmark Scheme 1}: {\bf Without PRA scheme}. This scheme corresponds to a MIMO radar with $N$ transmit and $M$ receive antennas without adjustable polarization, which is exactly the upper bound scheme in \cite{xu2023mimo}.
	\item {\bf Benchmark Scheme 2}: \textbf{Switchable PRA (SPRA) scheme.} This system consists of $N$ transmit and $M$ receive SPRAs, each connected to a single RF chain. Each antenna can switch between left- and right-handed circular polarization. The transmit and receive polarization vectors are ${\bm f} \in \left\{ \tfrac{1}{\sqrt{2}}[1,j]^T, \frac{1}{\sqrt{2}}[1,-j]^T \right\}$ and ${\bm e} \in \left\{ [1,j]^T, [1,-j]^T \right\}$, respectively.
	\item {\bf Benchmark Scheme 3}: \textbf{Circularly polarized antenna (CPA) scheme.} The polarization of all $N$ transmit and $M$ receive antennas is fixed to left-handed circular polarization. The transmit and receive polarization vectors are ${\bm f}= \frac{1}{\sqrt{2}}[1,j]^T$ and ${\bm e}= [1,j]^T$, respectively.
	\item{\bf Benchmark Scheme 4}: \textbf{Linearly polarized antenna (LPA) scheme.} All the $N$ transmit and $M$ receive antennas are fixed to vertical polarization. The transmit and receive polarization vectors are ${\bm f} = [1,0]^T$ and ${\bm e}= [1,0]^T$, respectively.
	\item {\bf Benchmark Scheme 5: Polarization-agile antennas (PAA) scheme.} The system employs $N$ transmit and $M$ receive PAAs, each capable of dynamically adjusting its polarization angle over $[0, 2\pi)$. The transmit and receive polarization vectors are defined as ${\bm f} = [\cos\zeta, \sin\zeta]^T$ and ${\bm e} = [\cos\eta, \sin\eta]^T$ \cite{castellanos2023linear}, respectively, where $\zeta$ and $\eta$ denote the corresponding polarization angles.
	\item {\bf Benchmark Scheme 6: Random phase scheme.} This scheme uses $N$ transmit and $M$ receive antennas, where each phase shifter applies an independent random phase shift uniformly drawn from $[0, 2\pi)$.
\end{itemize}	\vspace{-1mm}
\begin{figure}[!t]
	\centering
	\includegraphics[width=1\linewidth]{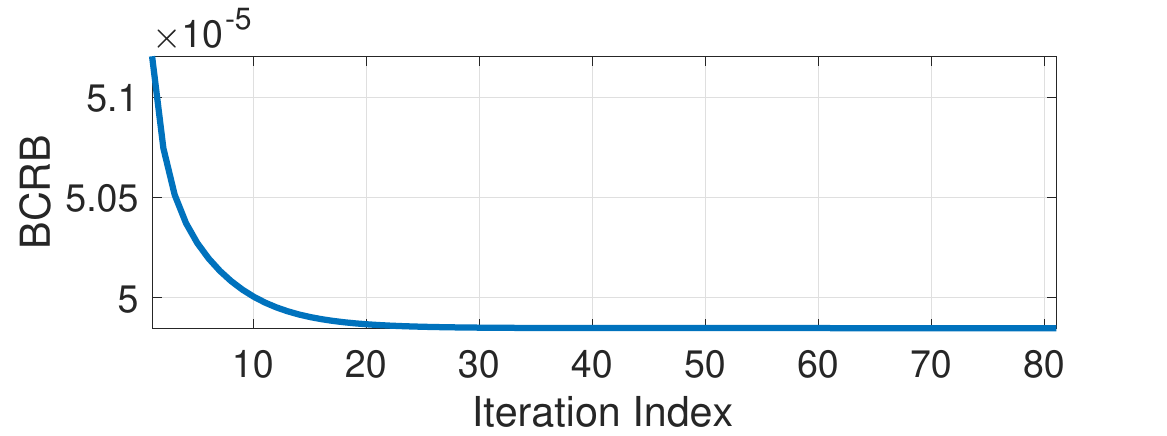}
			\vspace{-7mm}
	\caption{Convergence behavior of Algorithm \ref{algo1}.}
	\label{fig:Optimization_function_iter} 
		\vspace{-3mm}
\end{figure}

\begin{figure}[!ht]
	\centering
	\includegraphics[width=1\linewidth]{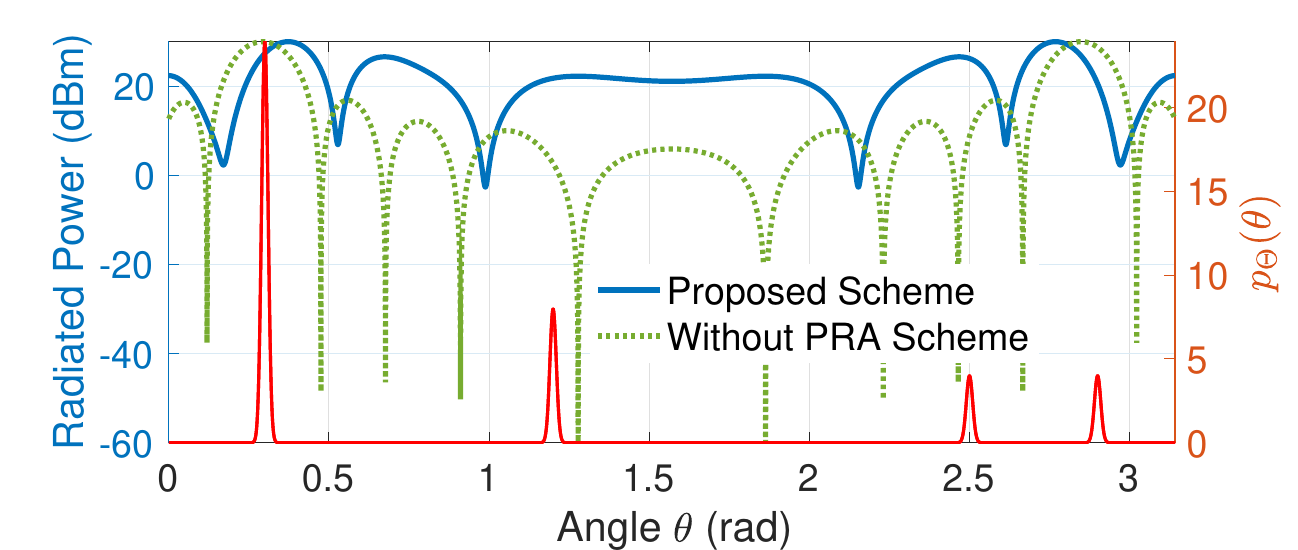}
	\vspace{-8mm}\caption{Radiated power pattern and $p_\Theta(\theta)$ over different angles.}
	\label{fig:Pattern} 
	\vspace{-5mm}
\end{figure}

\begin{figure}[!ht]
	\centering
	\includegraphics[width=1\linewidth]{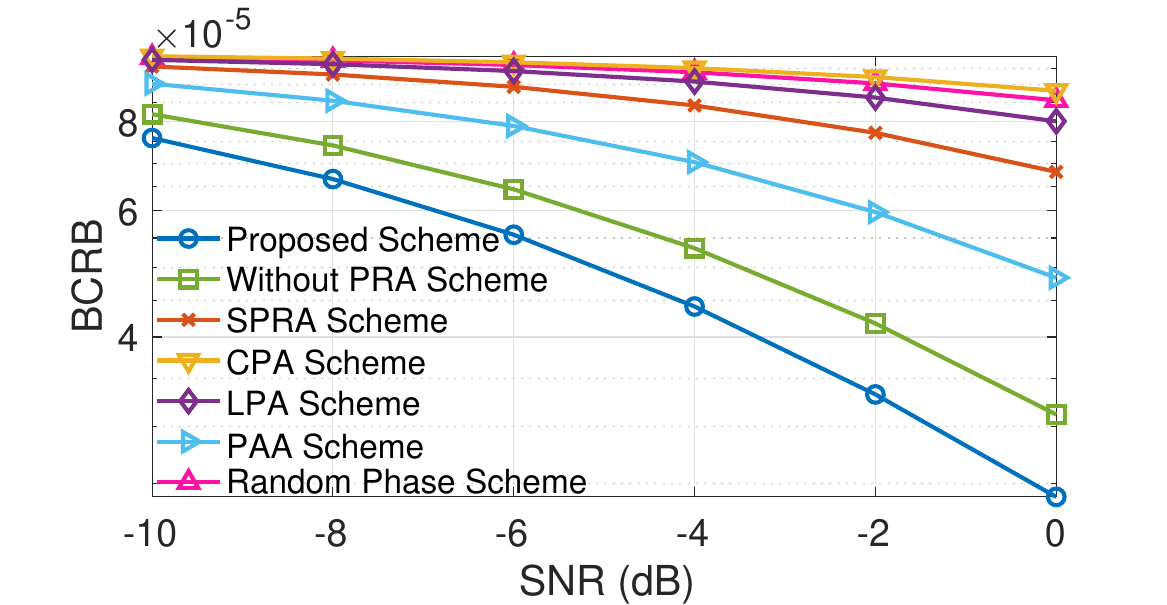}
	\vspace{-6mm}\caption{BCRB versus received SNR for proposed and benchmark schemes.}
	\label{fig:PCRB_SNR} 
	\vspace{-3mm}
\end{figure}

First, we illustrate in Fig. \ref{fig:Optimization_function_iter} the convergence behavior of the proposed AO-based algorithm. It is observed that the proposed algorithm converges monotonically within a few iterations.

Next, we show in Fig. \ref{fig:Pattern} the radiated power pattern with different transmit signal designs and the prior PDF of $\theta$ over different angles. It is observed that the proposed scheme exhibits superior performance across most angles with non-zero probability densities, compared with the Benchmark Scheme 1, particularly at those with moderate prior probabilities.

Finally,  in Fig.~\ref{fig:PCRB_SNR}, we evaluate the BCRB versus the received SNR $\frac{PL\gamma}{\sigma_\rms^2}$  achieved by different antenna polarization configurations. It is observed that the proposed PRA-aided MIMO radar consistently achieves lower BCRB compared to both conventional MIMO radar and other polarized MIMO radar schemes. This performance gain shows the advantages of continuous phase adjustment of the phase shifters and the effective phase optimization enabled by the proposed AO-based algorithm, highlighting the enhanced sensing capability of the proposed PRA-aided MIMO radar.
\vspace{-1mm}
\section{Conclusions}
This paper studied a novel PRA-aided MIMO radar system for sensing the unknown location parameter of a point target. With only prior distribution information about the parameter to be sensed, we derived the BCRB of the MSE, based on which we formulated the joint optimization problem of the transmit sample covariance matrix, the transmit phase shift vectors, and the receive phase shift vectors for the PRAs towards BCRB minimization. Despite the non-convexity of the problem, we devised an AO-based algorithm which iteratively obtains the optimal solution to one optimization variable in closed form, with the other variables being fixed at each time, thereby converging to at least a stationary point. The effectiveness of the proposed algorithm under the PRA-aided MIMO radar system was validated via numerical results.
\vspace{-1mm}

\bibliographystyle{IEEEtran}
\bibliography{IEEEref}
\end{document}